\documentclass[aoas,preprint]{imsart}

\RequirePackage[OT1]{fontenc}
\RequirePackage{amsthm,amsmath, amssymb}
\RequirePackage{natbib}
\RequirePackage[colorlinks,citecolor=blue,urlcolor=blue]{hyperref}
\usepackage{graphicx}
\usepackage{bbm}
\usepackage{array}
\usepackage{soul}
\newcolumntype{C}[1]{>{\centering\let\newline\\\arraybackslash\hspace{0pt}}m{#1}}

\arxiv{arXiv:1608.05655}

\startlocaldefs
\numberwithin{equation}{section}
\theoremstyle{plain}

\endlocaldefs

\newcommand{\bftheta}{{\boldsymbol \theta}}

\newcommand{\bfs}{{\bf s}}
\newcommand{\mcP}{\mathcal{P}}
\newcommand{\mcS}{\mathcal{S}}
\DeclareMathOperator{\Cov}{Cov}
\DeclareMathOperator{\BICM}{BICM}
\DeclareMathOperator{\AICM}{AICM}
\DeclareMathOperator{\CRPS}{CRPS}

\DeclareMathOperator{\R}{R}

\begin{document}

\begin{frontmatter}
\title{Nonstationary Spatial Prediction of Soil Organic Carbon:  Implications for Stock Assessment Decision Making}
\runtitle{Nonstationary Spatial Prediction of Soil Organic Carbon}

\begin{aug}
\author{\fnms{Mark D.} \snm{Risser}\thanksref{m1}\ead[label=e1]{mdrisser@lbl.gov}},
\author{\fnms{Catherine A.} \snm{Calder}\thanksref{m2}\ead[label=e2]{calder@stat.osu.edu}},
\author{\fnms{Veronica J.} \snm{Berrocal}\thanksref{m3}\ead[label=e3]{berrocal@umich.edu}},
\and
\author{\fnms{Candace} \snm{Berrett}\thanksref{m4}\ead[label=e4]{cberrett@stat.byu.edu}}

\runauthor{M. D. Risser et al.}

\affiliation{Lawrence Berkeley National Laboratory\thanksmark{m1}, The Ohio State University\thanksmark{m2}, University of Michigan\thanksmark{m3}, and Brigham Young University\thanksmark{m4}}

\address{Climate \& Ecosystem Sciences Division\\
Lawrence Berkeley National Laboratory\\
1 Cyclotron Road\\
Berkeley, CA 94720\\
\printead{e1}}

\address{Department of Statistics\\
The Ohio State University\\
429 Cockins Hall\\
1958 Neil Avenue\\
Columbus, OH 43210\\
\printead{e2}}

\address{Department of Biostatistics\\
University of Michigan\\
M4525 SPH II\\
1415 Washington Heights\\
Ann Arbor, MI 48109\\
\printead{e3}}

\address{Department of Statistics\\
Brigham Young University\\
216 TMCB\\
Provo, UT 84602\\
\printead{e4}}
\end{aug}

\begin{abstract}
The Rapid Carbon Assessment (RaCA) project was conducted by the US Department of Agriculture's National Resources Conservation Service between 2010-2012 in order to provide contemporaneous measurements of soil organic carbon (SOC) across the US.  Despite the broad extent of the RaCA data collection effort, direct observations of SOC are not available at the high spatial resolution needed for studying carbon storage in soil and its implications for important problems in climate science and agriculture.  As a result, there is a need for predicting SOC at spatial locations not included as part of the RaCA project. In this paper, we
compare spatial prediction of SOC using a subset of the RaCA data for a variety of statistical methods. We investigate the performance of methods with off-the-shelf software available (both stationary and nonstationary) as well as a novel nonstationary approach based on partitioning relevant spatially-varying covariate processes. Our new method addresses open questions regarding (1) how to partition the spatial domain for segmentation-based nonstationary methods, (2) incorporating  partially observed covariates into a spatial model, and (3) accounting for uncertainty in the partitioning. In applying the various statistical methods we find that there are minimal differences in out-of-sample criteria for this particular data set, however, there are major differences in maps of uncertainty in SOC predictions. We argue that the spatially-varying measures of prediction uncertainty produced by our new approach are valuable to decision makers, as they can be used to better benchmark mechanistic models, identify target areas for soil restoration projects, and inform carbon sequestration projects.
\end{abstract}


\end{frontmatter}

\section{Introduction} \label{secIntroduction}

Oceans, terrestrial systems, and the atmosphere form the three primary carbon reservoirs on the earth (\citealp{Batjes1996}). The amount of carbon in terrestrial ecosystems is nearly three times that of the atmosphere, and while its size is dwarfed by the ocean's carbon storage, terrestrial carbon is much more dynamic (\citealp{Batjes1996}). Soil organic carbon (SOC), a generic term for the carbon found in soil's organic matter, constitutes approximately two-thirds of the carbon in terrestrial ecosystems and is an important component in the earth's carbon cycle (\citealp{Nature}). The carbon in soil is in continuous interaction with the atmosphere via processes such as plant growth and decomposition (\citealp{Bliss2014}), and SOC helps mitigate the negative consequences of global changes in climate by sequestering carbon released into the atmosphere by fossil fuel combustion (\citealp{EcoApp}; \citealp{Nature}). In addition to its relevance to the earth's climate, SOC is important in forestry and agriculture, as organic matter contributes to soil fertility by helping retain moisture and supply plant nutrients (\citealp{Bliss2014}; \citealp{Post2000}). Furthermore, SOC is one of the soil properties used by hydrologists to better understand how precipitation is processed by different land surfaces and contributes to surface and ground water quality (\citealp{Bliss2014}). 

Because of the broad relevance of SOC to climate and agriculture, the National Resources Conservation Service (NRCS) initiated the Rapid Carbon Assessment (RaCA) project in 2010 (\citealp{RaCA_Data}).  The goal of the RaCA project was to collect measurements of the carbon content of soil across the conterminous United States at a single point in time.  Specifically, the project emphasized collecting spatially-referenced SOC  measurements or ``stocks", i.e., the amount of SOC in a volume (area and depth of soil) to produce ``statistically reliable quantitative estimates of amounts and distribution of carbon stocks for U.S. soils under various land covers'' (\citealp{RaCA_Data}).  Since the collection of SOC data is highly limited by time and cost constraints (\citealp{Sleutel2003}; \citealp{Goidts2007}), SOC is measured only at limited locations. Thus, there is a need for statistical methods to predict SOC concentration at unobserved locations using the RaCA data.

In this paper, we consider the problem of spatial prediction of SOC concentration based on RaCA measurements.  While geostatistical methods have previously been used for this purpose (albeit not for the RaCA data set; see \citealp{Simbahan2006}), these analyses have relied on an assumption of second-order stationarity in SOC over space. As we demonstrate in Section \ref{secData}, this assumption is inappropriate because SOC shows evidence of second-order nonstationary behavior on regional scales. However, existing methods (both stationary and nonstationary) with off-the-shelf software are inadequate for spatial prediction of SOC, as they yield either unrealistic prediction maps or a noninformative characterization of prediction error. Since it is well-documented that SOC is influenced by covariate information such as land use (\citealp{EcoApp}) and soil properties (\citealp{Mishra2009}), we propose a novel approach for spatial prediction of SOC that uses these spatially-referenced covariates to describe the second-order nonstationarity in SOC.
Existing approaches for 
covariate-driven nonstationary spatial modeling (\citealp{calder08};  \citealp{schmidt11}; \citealp{reich2011}; \citealp{ViannaNeto}; \citealp{spde13}; \citealp{Risser15}) require fully-observed spatial covariates, however, in this case the relevant covariates are not observed everywhere a prediction is desired.  To address this limitation, we propose a covariate-partitioning approach to nonstationary spatial modeling that uses spatially-referenced covariate information to divide the spatial domain into distinct ``segments." Once defined, these segments partition the domain such that every location is contained in exactly one of these segments, and the SOC process within each segment is assumed to be locally stationary conditional on a particular segmentation.  Since it is the segment membership that will be used in our statistical model, not the value of the covariate, spatial prediction does not require covariates to be observed at the prediction location.  

In spite of the fact that the SOC data display meaningful nonstationarities on regional scales, we acknowledge that quantitative evaluation criteria do not indicate a strong preference for our nonstationary statistical model, relative to approaches with off-the-shelf software. This is not surprising (see, e.g., \citealp{Fuglstad2015}); in any case, another important feature of our approach is that we are able to capture spatial variation in the uncertainty associated with SOC concentration predictions.  Unlike traditional geostatistical methods where, for a fixed set of monitoring sites, spatial variation in prediction uncertainty is driven primarily by variation in the geographical distribution of the data (e.g., uncertainty is greatest at locations far from the observations), our approach readily captures how covariates such as land use and soil properties impact the strength of spatial dependence in SOC, which directly informs the assessment of soil carbon stocks.

The paper proceeds as follows: in Section \ref{secData}, we introduce the RaCA data in more detail and conduct exploratory analyses, and in Section \ref{secCovtPar} we introduce relevant explanatory variables and their subsequent partitioning. Section \ref{secModel} outlines our covariate-partitioning approach for nonstationary spatial prediction of SOC, and in Section \ref{secResults} we present the results of our fitted model, predictions, and implications for carbon stock assessment. Section \ref{secDiscussion} concludes the paper.

\section{RaCA data and exploratory analyses} \label{secData}

\begin{figure}[!t]
\begin{center}
\includegraphics[width=0.8\textwidth]{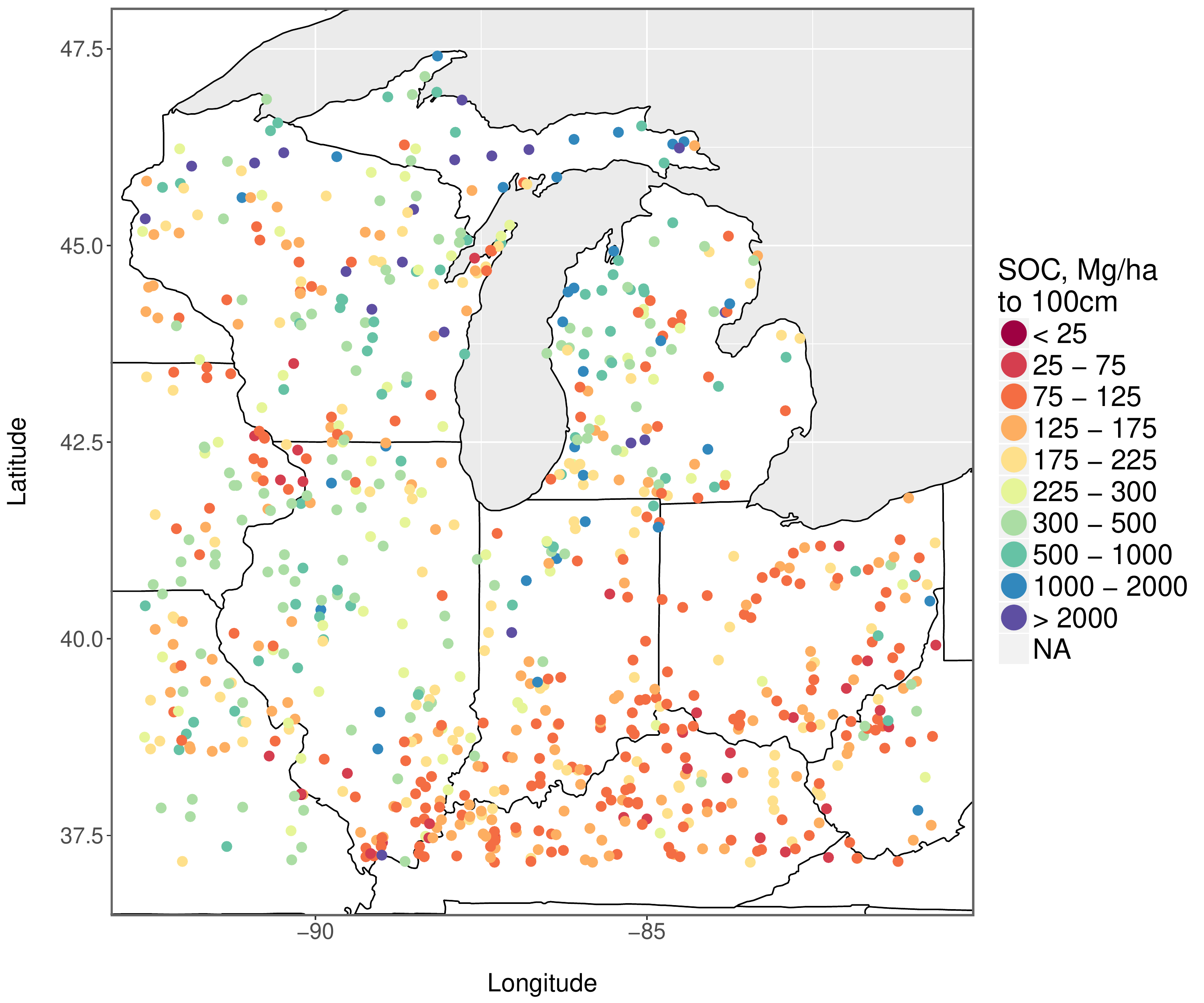}
\caption{Great Lakes subset of the soil organic carbon stocks data set. }
\label{SOCsubset}
\end{center}
\end{figure}

The RaCA data and additional variables such as land use-land cover (LULC) classes, soil series, and soil moisture are available through the {\tt soilDB} package in {\tt R} (\citealp{R_soilDB}), and a summary report of the sampling methods and data description is provided in \cite{RaCA_Data}. While measurements of SOC stocks are available for each site to depths of five, ten, twenty, thirty, fifty, and one hundred centimeters (in Mg C ha$^{-1}$), we use the one hundred centimeter depth measurement since our focus is on estimating total SOC, not a soil depth profile (e.g., \citealp{Minasny2006}; \citealp{Mishra2009}).  Our study focuses on a subset of the RaCA SOC measurements collected in the Great Lakes region of the midwestern United States, shown in Figure \ref{SOCsubset}, which contains $790$ observations. 

\subsection{Variogram analysis}

\begin{figure}[!t]
\begin{center}
\includegraphics[trim={65 0 93 0mm}, clip, width=\textwidth]{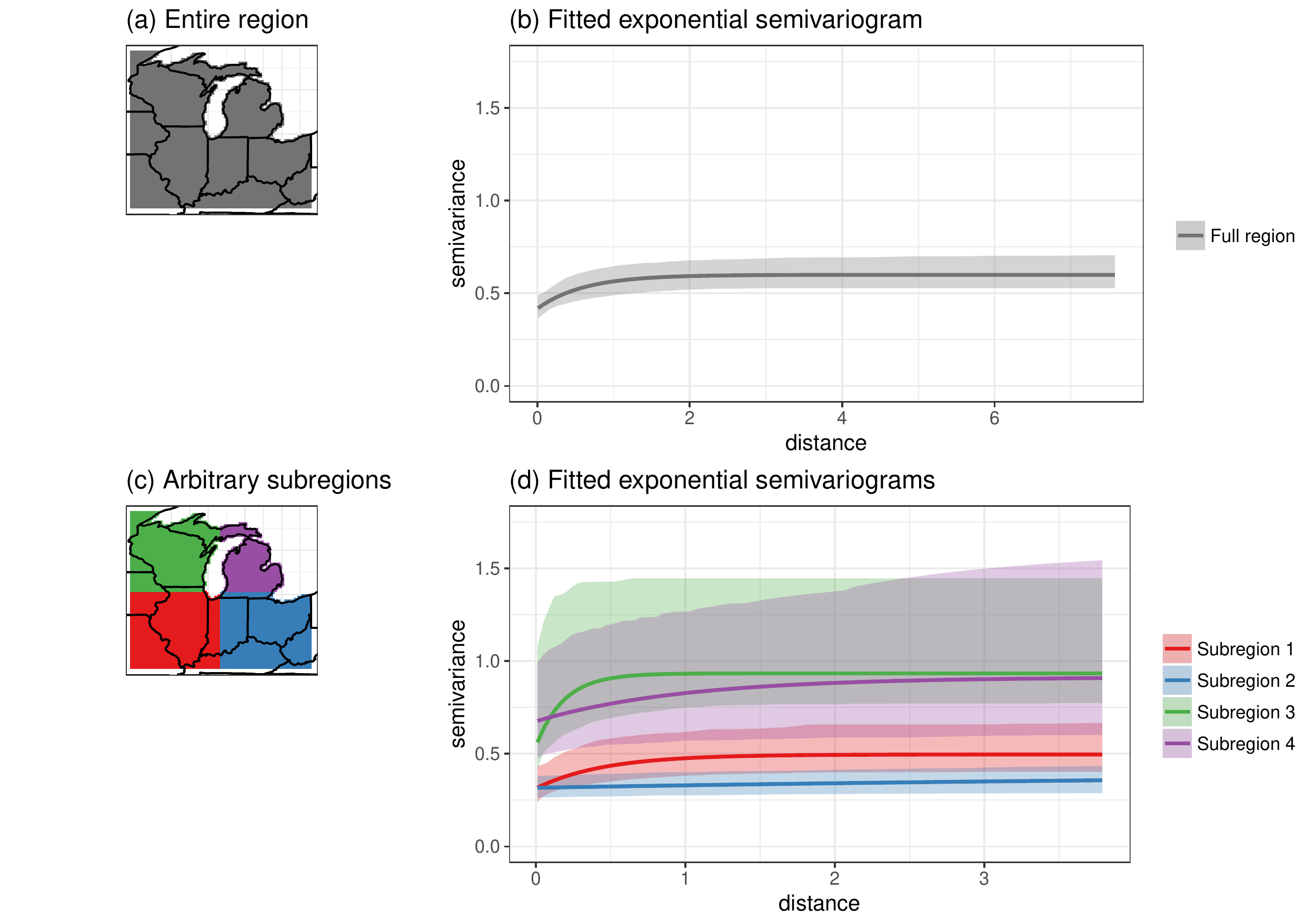}
\caption{Fitted exponential semivariograms for the entire Great Lakes region (top) and for four arbitrary subregions (bottom), with 95\% confidence bands.}
\label{fullVario}
\end{center}
\end{figure}

As discussed in Section \ref{secIntroduction}, accurate estimation of the spatial distribution of SOC is hindered by costly and time-consuming data collection, and we are thus motivated to consider spatial statistical methods to predict SOC concentration at unobserved locations. Focusing on the Great Lakes region, we begin with a simple variogram analysis of the SOC data, first calculating the empirical semivariogram and corresponding fitted exponential semivariogram for the entire region. Here and throughout the remainder of the paper, the SOC data is transformed to the $\log$ scale; for the variogram analyses, we use the ordinary least squares (OLS) residuals from a regression of log SOC on latitude, longitude, and the longitude/latitude interaction. Exploratory analysis indicates that the exponential correlation model fits the data well; the fitted exponential semivariogram is shown in Figure \ref{fullVario}(b) with 95\% confidence band (using the parametric bootstrap).

A simple way to assess the presence of second-order nonstationary behavior in a spatial data set is to split up the spatial domain into subregions and conduct a variogram analysis separately for each subregion. Arbitrarily dividing the Great Lakes region into four parts, we fit an exponential semivariogram to the OLS residuals in each subregion. The subregion-specific semivariograms {with 95\% confidence bands} (again using the parametric bootstrap) are shown in Figure \ref{fullVario}(d): the fitted semivariograms indicate that the subregions display quite different spatial dependence patterns. These differences are significant, as indicated by the non-overlapping uncertainty bands, which motivates a nonstationary spatial model for the SOC data where the spatial dependence properties vary over the spatial domain.  

\subsection{Off-the-shelf spatial prediction} \label{subsecOffshelf}
{Various packages in \texttt{R} are readily available for the analysis of spatial data, some developed for stationary processes and some accommodating nonstationary ones. Here, we summarize and report on the prediction results obtained by applying different models with off-the-shelf software to our SOC data.} All models were fitted using 20,000 MCMC iterations with the first 10,000 discarded as burn-in and with no thinning of the chain. All the spatial models use an exponential correlation structure, {as this was deemed to provide a good representation of the dependence structure in the data, and all use uninformative, most times uniform, priors on model parameters.}

\vskip2ex
\noindent {\textit{Bayesian additive regression trees}}. Bayesian additive regression trees (BART; \citealp{chipman:etal:2010}) is a Bayesian sum-of-trees model that has been shown to perform well in a variety of settings, particularly for prediction \citep[see, for example,][]{bonato:etal:2011, ding:etal:2012, green:kern:2012}. Because of its predictive skill, {we spatially predict (log) SOC using BART as implemented in the} \texttt{BayesTree} package for \texttt{R}. In {applying} BART {to the SOC data, we use} latitude and longitude  as covariates for the mean function. 

\vskip2ex
\noindent {\textit{Treed Gaussian process}}. {Related to BART is the non-stationary treed Gaussian process (TGP) model of \cite{GramacyLee} in which tree partitions are defined by segmenting the coordinate axes, and partitioning uncertainty is accounted for through a model averaging approach (\citealp{BMA1999}). The TGP model is implemented in the {\tt tgp} package for {\tt R} (\citealp{tgp}), which we use with all the defaults on (log) SOC. As in BART, latitude and longitude are used as main effects in the mean function.}

\vskip2ex
\noindent {\textit{Gaussian predictive process}}. {Given the moderately large dimension of the RaCA dataset}, {we also consider} an off-the-shelf implementation of the Gaussian predictive process (PP) model of \cite{Banerjee2008}, which is specifically designed to account for large spatial data sets.  The PP works by constructing an approximation to a Gaussian process  {by} projecting realizations of the process of interest onto a lower dimensional space spanned by the predictive process knots, thereby reducing the computational burden. {By construction, the PP is technically non-stationary, and the PP model provides a non-stationary approximation to any stationary covariance function.} We fit a PP {model to our SOC data using the {\tt spBayes} package for {\tt R} (\citealp{Finley2007}; \citealp{Finley2013}) specifying $103$ knots and a constant mean.} 

\vskip2ex
\noindent {\textit{Bayesian stationary Gaussian process}}. 
{The PP model reduces to a Bayesian stationary Gaussian process model if the predictive process knots are taken to be exactly the observation locations. Using the {\tt spBayes} package and the {\tt spLM} function with default prior settings, we generate spatial predictions of log SOC using a traditional Bayesian stationary (isotropic) spatial Gaussian process model with mean (log) SOC specified as a linear function of longitude and latitude.}

\begin{table}[!t]
\caption{A summary of the models fit to the SOC data. (GP indicates a Gaussian process; lat/lon refers to latitude/longitude.) All spatial models use an exponential correlation function. Computational times correspond to fitting the model to the full data set ($n = 790$).}
\begin{center}
\begin{tabular}{|p{1.15cm}|p{3.5cm}|p{1.5cm}|p{1.6cm} | p{2.35cm}|}
\hline
\textbf{Label} 	& \textbf{Details} 				& \textbf{Mean function} 		& \textbf{\texttt{R} package} & \textbf{Computational time}		\\ \hline\hline 
BART$^\ddagger$		& Bayesian additive regression trees & Lat/lon as inputs		& \texttt{BayesTree}	& 12.4 minutes$^\dagger$	 	\\ \hline
TGP$^\ddagger$			& Bayesian treed GP			& Lat/lon for inputs and mean	& \texttt{tgp} & 13.1 minutes$^\dagger$	 	\\ \hline 
PP$^\ddagger$			& Bayesian Predictive Process (r = 103 knots) & Constant	& \texttt{spBayes}		& 2.4 minutes$^\dagger$	 	\\ \hline
SGP$^\ddagger$		& Bayesian stationary GP 			& Lat/lon				& \texttt{spBayes} & 19.4 minutes$^\dagger$		\\ \hline \hline
NSGP		 & Bayesian nonstationary GP   		& Constant				& n/a & 7.0 (max), 3.4 (avg.) minutes$^\dagger$	 	\\ \hline
\end{tabular}
\end{center}
\label{modelComp}
\begin{flushleft}
\vskip-2ex
{\scriptsize $^\dagger$Time given for an Intel Core i7 3.1 GHz machine (16 GB memory) for 20000 MCMC iterations.}\\
{\scriptsize $^\ddagger$Methods using available off-the-shelf software.}\\
\end{flushleft}
\end{table}%

\begin{figure}[!t]
\begin{center}
\includegraphics[width=\textwidth]{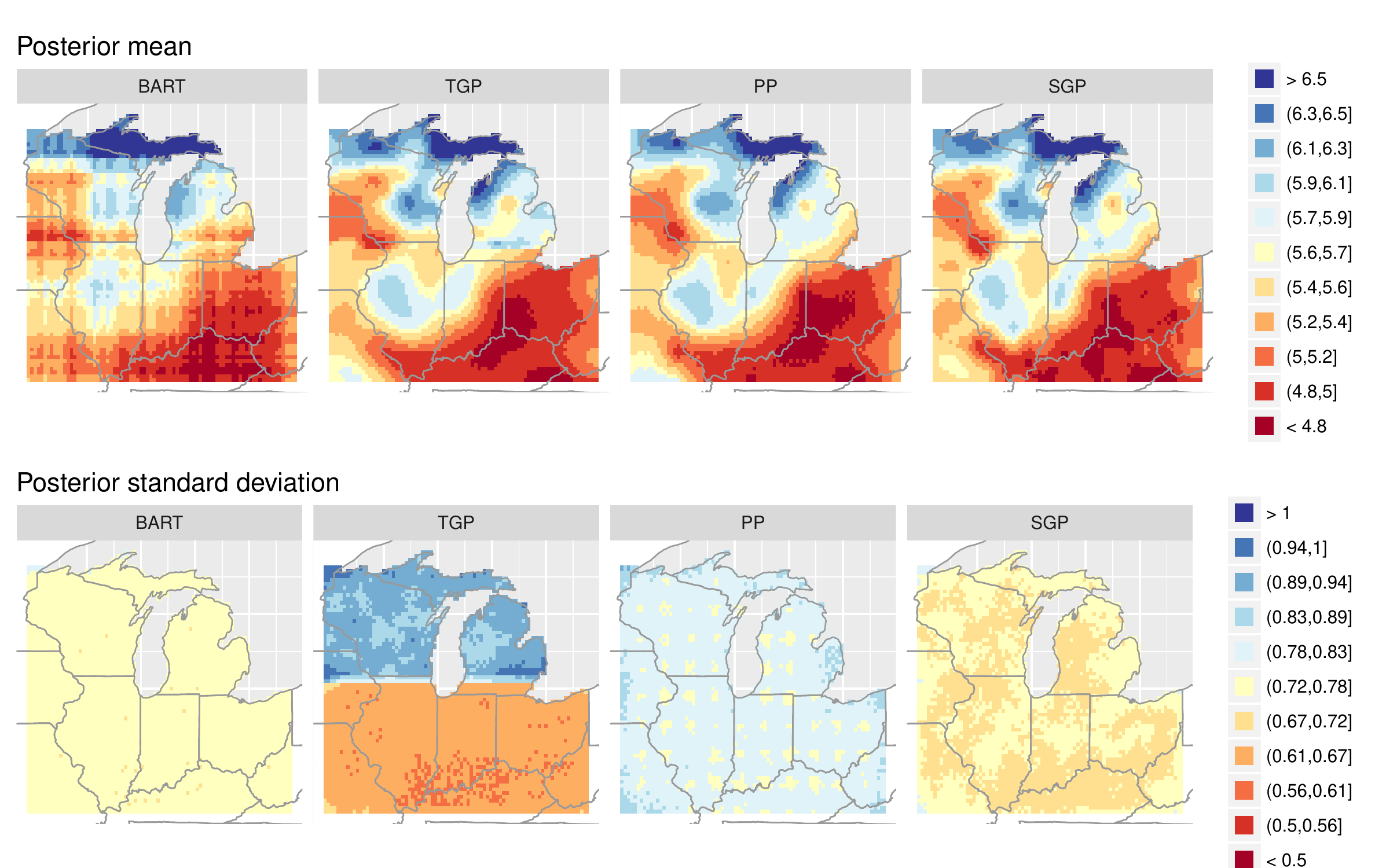}
\caption{Posterior mean predictions (top) and corresponding posterior standard deviations (bottom) for log SOC in the Great Lakes region, using the models BART (left), TGP (left-center), PP (right-center), and SGP (right). (Note: both color bars have the same limits as in Figure \ref{maPreds}.)}
\label{otherPreds}
\end{center}
\end{figure}

\vskip2ex
\noindent {Rows 2-5 in Table \ref{modelComp} summarize the different models applied to the (log) SOC data, with details on the mean function specification, the \texttt{R} package used to implement the model, and the computational time that it takes to fit each model. Surfaces of (log) SOC obtained by generating predictions on a fine grid covering the entire spatial domain using each of the aforementioned methods are presented in Figure \ref{otherPreds}. Specifically, from left to right, the panels show the posterior mean (top) and standard deviation (bottom) of the predictions using BART, TGP, SGP, and PP. Clearly, none of these models appear to be appropriate for the data: 
BART does not yield scientifically meaningful predictions with the artificial horizontal and vertical lines, and while the three other spatial models produce prediction maps that are generally smooth they are characterized by uncertainty prediction maps that are either unrealistic  for an environmental process like log SOC (see TGP), or uninformative since they are almost constant in space.}

{To address the limitations of these off-the-shelf methods, and to model the globally non-stationary but locally stationary nature of the SOC data highlighted in Section~\ref{secData}, in Sections \ref{secCovtPar} and \ref{secModel} we move on to introduce a novel segmentation-based model for non-stationary spatial processes with partitions informed by covariates.}

\section{Covariate-driven partitioning} \label{secCovtPar}

\sloppypar{
Returning to the exploratory analysis in Section \ref{secData}, it is clear that partitioning the domain captures {the} nonstationary behavior in SOC for the Great Lakes region. However, there are three problems associated with the subregion-specific variogram analyses. First, the arbitrary partitioning of our domain as in Figure \ref{fullVario} is not scientifically meaningful: in other words, it provides no way of understanding why the different subregions exhibit nonstationarities. Second, generating predictions of SOC based only on separately or independently fitted variograms for each subregion does not comprise an appropriate spatial model, because no information is shared across the subregions. Third, the fitted semivariogram estimates are likely sensitive to the specific partition used in Figure \ref{fullVario}; in other words, we might want to account for uncertainty in the partition itself.}

\begin{figure}[!t]
\begin{center}
\includegraphics[trim={0 0 40 0mm}, clip, width=\textwidth]{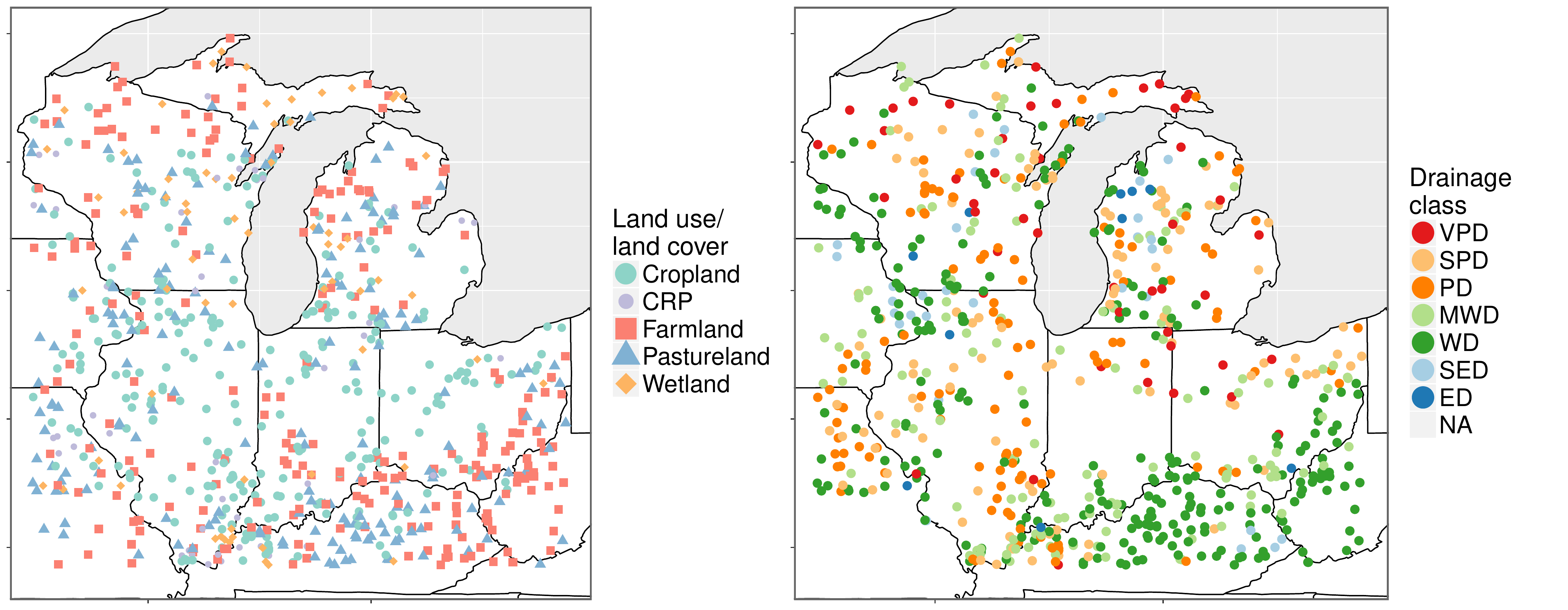}
\caption{The Great Lakes region subset of the Rapid Carbon Assessment (RaCA) land use-land cover classes (left) and drainage classes. For land use, CRP refers to a Conservation Reserve Program cropland site. For drainage class, the labels are as follows: VPD = very poorly drained, SPD = somewhat poorly drained, PD = poorly drained, MWD = moderately well drained, WD = well drained, SED = somewhat excessively drained, ED = excessively drained. }
\label{LU_DC}
\end{center}
\end{figure}

A solution to the first problem is to partition the spatial domain based on {covariate information}. As discussed in Section \ref{secIntroduction}, there are well-documented relationships between SOC and a number of covariate variables, for example, land use-land cover (LULC) class and drainage class. Both of these categorical variables are available in the {\tt soilDB} package in {\tt R} (\citealp{R_soilDB}).  According to \cite{RaCA_Data}, the LULC classes were developed specifically for the RaCA project to correspond to the classes and definitions of the Natural Resources Inventory. RaCA designated five specific LULC classes (four of which are represented in the Great Lakes region subset), namely cropland, farmland, pastureland, rangeland, and wetland, with one additional category for any cropland site that was also known to correspond to a Conservation Reserve Program (CRP). The drainage class variable refers to ``the frequency and duration of wet periods under conditions similar to those under which the soil developed. Alteration of the water regime by man, either through drainage or irrigation, is not a consideration unless the alterations have significantly changed the morphology of the soil'' (NRCS Soil Survey Manual, chapter 3). Plots of the subsetted LULC and drainage class variables are shown in Figure \ref{LU_DC}. 

Given the known relationships between SOC and these variables as well as the rich literature on covariate-driven nonstationary modeling, we are motivated to use these covariates to describe the second-order nonstationary behavior exhibited by SOC over the Great Lakes region. In other words, our hypothesis is that both the first- and second-order properties of SOC might be similar in areas where the covariates are homogeneous. {H}owever, measurements of LULC and drainage class are not available for every prediction location of interest, {which render} the nonstationary models of \cite{reich2011}, \cite{ViannaNeto}, \cite{spde13}, and \cite{Risser15} {unusable in this situation}.

{Hence, here we propose} to partition the spatial domain based on the multivariate spatial distribution of LULC and drainage class. This provides scientifically meaningful partitioning, and also accounts for the fact that we are dealing with incompletely observed covariates. {Having defined a partitioning of the spatial domain, we can use an} approach {similar to one} outlined in \cite{fuentes2001} (see Section \ref{secModel}) {to define a globally non-stationary, locally-stationary spatial process}. Finally,  to both account for uncertainty in the partitioning \textit{and} introduce spatial dependence in SOC across subregions{, we propose to use a Bayesian model averaging framework (\citealp{BMA1999})}.

There are a variety of methods in the statistics literature for obtaining partitions of a multivariate space (here, geographic space, i.e., latitude and longitude) based on multivariate inputs (here, LULC and drainage class). For this paper, we use multivariate cluster-wise regression, also called multivariate latent class regression (see, e.g., \citealp{Leisch2004}; \citealp{Muller2011}). Before describing this approach, we establish some terminology: we define a \textit{partition} of the spatial domain $\mathcal{D}$ as the assignment of each location in the geographical space to a particular subregion. In a partition, each location is assigned to exactly one subregion and collectively the subregions comprise the entire spatial domain. Also, we define a \textit{segment} as an individual subregion: therefore, a partition is made up of a set of segments. We use $\mcP$ to denote a generic partition, made up of segments $\{ \mcS_1, \dots, \mcS_K \}$. By definition, $\mcP = \cup_k \mcS_k$ and $\mcS_k \cap \mcS_{k'} = \varnothing$ for $k \neq k'$.

\begin{figure}[!t]
\begin{center}
\includegraphics[width=\textwidth]{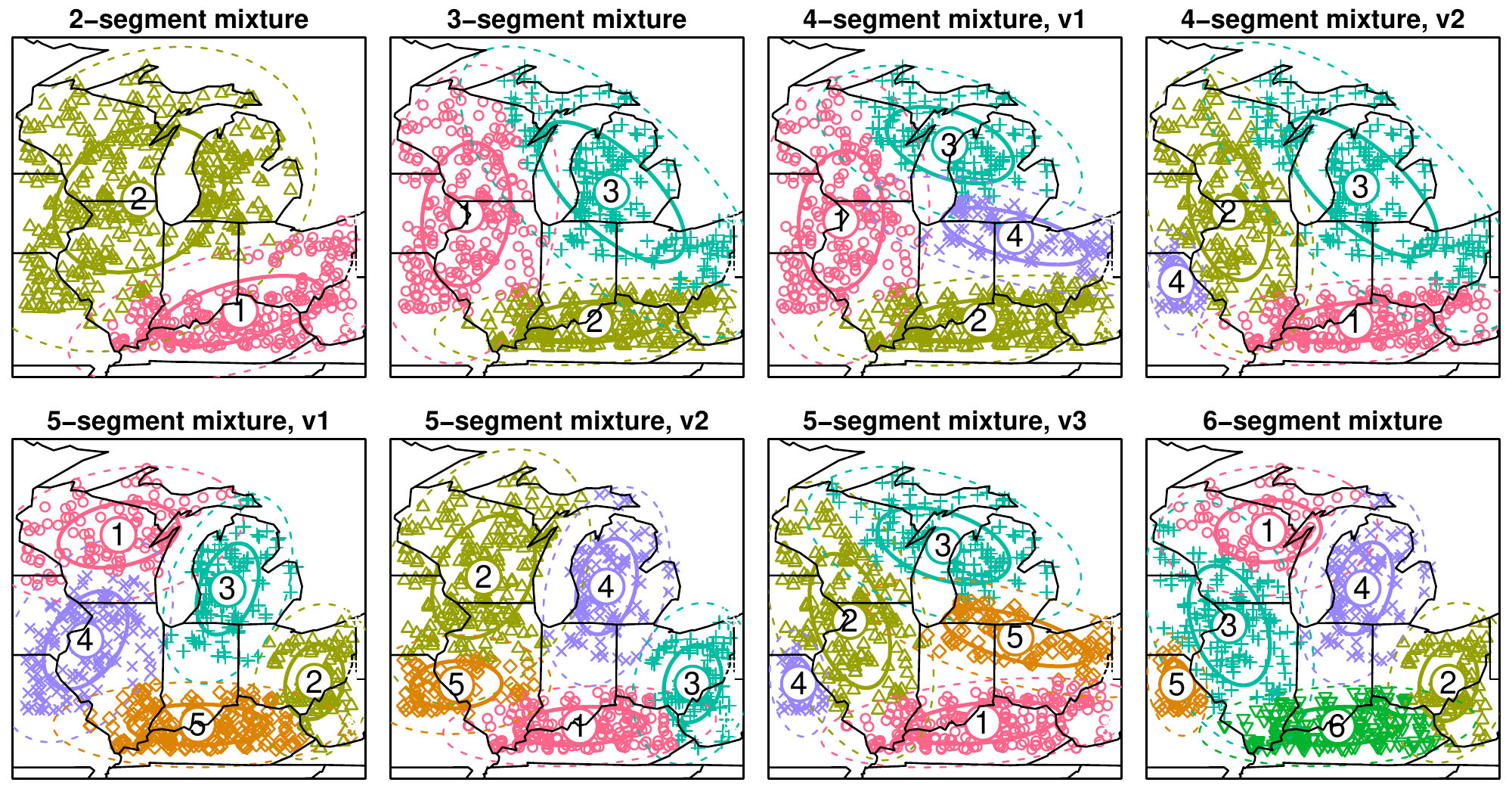}
\caption{Finite mixture models used to generate the candidate partitions. The plotted points represent locations with a non-missing observation of both the drainage class and land use variables.}
\label{mixtures}
\end{center}
\end{figure}

It is trivial to note that determining the segments of a partition $\mathcal{P}$ of  the spatial domain $\mathcal{D}$ is equivalent to clustering the geographical coordinates of points within $\mathcal{D}$. In particular, if the segments of the partition are to be characterized by the fact that the covariate processes are similar within each segment, then the clustering mechanism ought to be informed by the covariates.  \cite{Leisch2004} and \cite{Muller2011} propose two different approaches to achieve covariate-informed clustering: the first approach falls in the model-based clustering category, and the second falls in the class of product partition models (thus, specified in a Bayesian nonparametric framework). For computational simplicity, we follow \cite{Leisch2004} and use a model-based clustering approach. Working with a two-dimensional spatial domain (i.e., latitude and longitude of locations) we model the bivariate vector of coordinates $(s_1, s_2)$ for each point $\mathbf{s}\in \mathcal{D}$ as arising from a mixture of bivariate normal distributions where the mean and the covariance matrix are mixture-component specific. In other words, given a pre-specified and finite number $K$ of clusters, we assume
\begin{equation}
\bfs = \left( s_1, s_2 \right)  \sim \sum_{k=1}^{K} \pi_k(\mathbf{s}) \cdot N_{2} (\bfs; \mathbf{m}_k, \mathbf{D}_k),
\label{eq:clusters}
\end{equation}
{where $N_{d} (x; m, V)$ is the $d$-variate Gaussian density with mean $m$ and covariance $V$ evaluated at $x$.}
The \texttt{flexmix} package in \texttt{R} allows us to fit this class of models for fixed $K$ using the EM algorithm. Taking the observation locations of the two spatial covariates (LULC and drainage class) as data for model (\ref{eq:clusters}) and using the \texttt{flexmix} package with $K$ ranging from 2 to {6}, we obtain multiple potential partitions of the observation locations of the covariate processes, shown in Figure \ref{mixtures}. The mixture for a specified number of segments is non-unique (in other words, the EM algorithm converges to several different local modes), and after re-starting the algorithm several times for each $K$ we selected the best mixture(s) based on EM convergence and the log-likelihood.

However, note that Figure \ref{mixtures} only provides partitions for locations where we have a measurement of the spatial covariates. As will be seen in Section \ref{secModel}, we require partitions of the locations where we have SOC measurements (which differ from the covariate measurement locations) as well as all locations on a fine grid for generating predictions. When fitting model (\ref{eq:clusters}) to the latitude and longitude coordinates of the covariate locations using the package \texttt{flexmix}, we obtain as a byproduct the {bivariate Gaussian densities from (\ref{eq:clusters}), which can be defined for any location $\mathbf{s}\in \mathcal{D}$. Based on the $K$ segment probabilities, we assign each location to a segment by taking the maximum of the bivariate Gaussian densities, i.e., the segment for location $\bfs$ is defined as $\max_k \{ N_{2} (\bfs; \mathbf{m}_k, \mathbf{D}_k) \}$. Thus the eight mixtures in Figure \ref{mixtures} yield eight partitions of the spatial domain (not shown), denoted $\{ \mcP_j: j = 1, \dots, 8 \}$, with segments denoted $\{ \mcS_{jk} : k = 1, \dots, K_j\}$.}


\section{A partition-based nonstationary spatial Gaussian process model} \label{secModel}

The covariate partitions $\{ \mathcal{P}_j: j = 1, \dots, 8 \}$ {defined in  Section~\ref{secCovtPar}} can be used to model {both} first- and second-order nonstationarities in SOC as follows.  Let $Z(\cdot)$ represent observed log SOC, where we model $Z(\cdot)$ as a spatial stochastic process defined 
for all $\bfs\in\mathcal{D}$ (here, $\mathcal{D}$ denotes the Great Lakes region of the United States). For all ${\bf s} \in \mathcal{D}$, let
\begin{equation} \label{model}
Z({\bf s}) = \mu({\bf s}) + Y({\bf s}) + \varepsilon({\bf s}), 
\end{equation}
where $E[Z({\bf s})] = \mu({\bf s})$ is a deterministic mean function, $Y(\cdot)$ is a mean-zero latent spatial Gaussian process, and $\varepsilon(\cdot)$ is an error process that is assumed to be independent of $Y(\cdot)$. We observe the value of $Z(\cdot)$ at a fixed, finite set of locations $\{ \bfs_1, \bfs_2, \dots, \bfs_n \} \in \mathcal{D}$ (see Figure \ref{SOCsubset}) and wish to use these observations to learn about the underlying processes and generate predictions at unobserved locations. 

In Section \ref{subsecCond}, we outline a statistical model for log SOC conditional on a single partition $\mcP_j$, and in Section \ref{subsecMCMC} we outline the model fitting for each conditional model. Then, in Section \ref{subsecBMA} we describe a model-averaging approach to posterior prediction that incorporates all candidate partitions.

\subsection{Conditional model specification} \label{subsecCond}

Conditional on the $K_j$ segments $\{ \mcS_{jk} : k = 1, \dots, K_j\}$ of partition $\mcP_j$, we first model $Y(\cdot)$ as a mixture of stationary processes (\citealp{fuentes2001}), i.e., 
\begin{equation}\label{discretesum}
Y(\bfs) = \sum_{k=1}^{K_j}  w_{jk}(\bfs)\widetilde{Y}_{jk}(\bfs),
\end{equation}
where the $\widetilde{Y}_{jk}(\cdot)$ are orthogonal 
and stationary, and $w_{jk}(\cdot)$ is a positive kernel weight function such that $w_{jk}(\bfs) \geq 0$ and $\sum_{k=1}^{K_j }[w_{jk}(\bfs)]^2 = 1$ for all $\bfs \in \mathcal{D}$ (following \citealp{reich2011}). Define the covariance function  of each $\widetilde{Y}_{jk}(\cdot)$ to be $\widetilde{C}_{jk}$; then, the covariance function of $Y(\cdot)$ is
\begin{equation} \label{discSumCF}
\Cov\big(Y(\bfs), Y(\bfs')\big) \equiv C(\bfs, \bfs') = \sum_{k=1}^{K_j} w_{jk}(\bfs) w_{jk}(\bfs') \widetilde{C}_{jk}(\bfs - \bfs'),
\end{equation}
which is a valid nonstationary covariance function. For each $\widetilde{C}_{jk}$, we use an anisotropic version of the parametric Mat\'ern model
\begin{equation} \label{maternMod}
\widetilde{C}_{jk}(\bfs - \bfs'; \bftheta_{jk}) = \frac{ \sigma_{jk}^2 }{\Gamma(\nu_{jk})2^{\nu_{jk} - 1}} \left[ \sqrt{Q_{jk}(\bfs-\bfs')} \right]^{\nu_{jk}} \mathcal{B}_{\nu_{jk}} \left( \sqrt{Q_{jk}(\bfs-\bfs')}  \right).
\end{equation} 
In (\ref{maternMod}), $\mathcal{B}_{\nu_{jk}}(\cdot)$ denotes the modified Bessel function of the third kind of order $\nu_{jk}$,
$Q_{jk}(\bfs - \bfs') = ||{\bf \Sigma}^{-1/2}_{jk} (\bfs - \bfs')||^2$ is a squared Mahalanobis distance with anisotropy matrix ${\bf \Sigma}_{jk}$ parameterized according to its spectral decomposition, i.e. (for $d=2$),
\begin{equation} \label{anisoMatrix}
{\bf \Sigma}_{jk} = \left[ \begin{array}{cc} \cos(\eta_{jk}) & - \sin(\eta_{jk})  \\ \sin(\eta_{jk}) & \cos(\eta_{jk})  \end{array} \right] \left[ \begin{array}{cc} \phi^{(1)}_{jk} & 0  \\ 0 & \phi^{(2)}_{jk}  \end{array} \right] \left[ \begin{array}{cc} \cos(\eta_{jk}) & \sin(\eta_{jk})  \\ -\sin(\eta_{jk}) & \cos(\eta_{jk})  \end{array} \right],
\end{equation}
and $\bftheta_{jk} = (\sigma^2_{jk}, \nu_{jk}, \phi^{(1)}_{jk}, \phi^{(2)}_{jk}, \eta_{jk})$ is a vector of parameters that control the variance, smoothness, and anisotropy of $\widetilde{Y}_{jk}(\cdot)$. In the anisotropy matrices (\ref{anisoMatrix}), $\phi^{(1)}_{jk}$ and $\phi^{(2)}_{jk}$ represent directional ``ranges'' (i.e., inverse decay parameters) and $\eta_{jk}$ represents an angle of rotation; these parameters allow for locally elliptical correlation patterns. 

In general, we might propose a similar model for the mean behavior $\mu(\cdot)$, i.e., $\mu(\bfs) = \sum_{k=1}^{K_j}  w_{jk}(\bfs)\widetilde{\mu}_{jk}(\bfs)$, where $\widetilde{\mu}_{jk}(\bfs)$ can accommodate any mean structure, including (in the case of a linear mean) fully observed covariates and segment-specific intercepts. For log SOC, we considered a variety of segment-specific mean functions involving latitude and longitude, but found that using  a constant mean across all segments performed as well as a model with a global mean latitude and longitude coefficients. 
{Furthermore, a model with constant spatial mean performed as well as a model with a different intercept in each segment.}
Thus, we set $\widetilde{\mu}_{jk}(\bfs) \equiv \mu_j$ for all $k$.

To complete the specification of our model (\ref{model}), we suppose that the error process $\varepsilon({\bf s})$ is spatially independent and Gaussian. Similar to (\ref{discretesum}), we model $\varepsilon(\bfs) = \sum_{k=1}^{K_j}  w_{jk}(\bfs)\widetilde{\varepsilon}_{jk}(\bfs)$, where each $\widetilde{\varepsilon}_{jk}(\bfs) \stackrel{\text{iid}}{\sim} N(0, \tau^2_{jk})$. Thus, we expand the $\bftheta_{jk}$ vector to include the variance $\tau^2_{jk}$; collect all of the variance/covariance parameters across segments into a single vector 
\[
\bftheta^{(j)} = \{ \bftheta_{j1}, \dots, \bftheta_{jK_j} \}.
\]

In this paper, the weight functions $w_{jk}(\cdot)$ for partition $\mcP_j$ are chosen to be indicator functions for segment $\mcS_{jk}$, i.e., $w_{jk}(\bfs) = \mathbbm{1}(\bfs \in \mcS_{jk})$ (following, e.g., \citealp{GramacyLee}). In this case, for partition $\mcP_j$, the process $Y(\cdot)$ and therefore $Z(\cdot)$ is now locally stationary within each $\mcS_{jk}$ and independent across the $\mcS_{jk}$, conditional on $\mcP_j$. For the random observed vector ${\bf Z} \equiv ( Z(\bfs_1), \dots, Z(\bfs_n) )^\top$, the indicator weight function implies a conditional likelihood for ${\bf Z}$,
\begin{equation} \label{likelihood}
p({\bf Z}| \mu_j, \bftheta^{(j)}, \mcP_j) \propto \prod_{k=1}^{K_j} \Big|{\bf V}_{jk} + {\bf \Omega}_{jk}\Big|^{-1/2}  \hskip27ex
\end{equation}
\[ 
\hskip15ex \times \exp\left\{ -\frac{1}{2} ({\bf Z}_{jk} - \mu_j{\bf 1}_{jk})^\top ({\bf V}_{jk} + {\bf \Omega}_{jk})^{-1} ({\bf Z}_{jk} - \mu_j{\bf 1}_{jk}) \right\},
\]
which is the product of segment-specific multivariate Gaussian likelihoods. In (\ref{likelihood}), the ``$jk$'' subscript partitions each term into its partition- and segment-specific components, e.g., ${\bf Z}_{jk} = \{ Z(\bfs_i): \bfs_i \in \mcS_{jk}\}$; ${\bf V}_{jk} = \tau^2_{jk} {\bf I}$ captures a partition-/segment-specific measurement error variance; the elements of ${\bf \Omega}_{jk}$ come from $\widetilde{C}_{jk}$. Note that the likelihood for ${\bf Z}$ in (\ref{likelihood}) is \textit{conditional} on $\mcP_j$. As a result, the partition controls the second-order properties of $Z(\cdot)$ by determining independent regions of local stationarity. As mentioned previously, the partition could also specify first-order properties, although for SOC we have not used this property.

Still conditional on $\mcP_j$, we factor the prior distribution as $p(\mu_j, \bftheta^{(j)} |\mcP_j) = p(\mu_j |\mcP_j) \times p(\bftheta^{(j)} |\mcP_j)$. The prior on $\mu_j$ is proper but noninformative and conjugate for the likelihood (\ref{likelihood}), i.e., $p(\mu_j |\mcP_j) = N({ 0}, 100^2)$.
The prior for $\bftheta^{(j)}$ is conditional on the partition since $\mcP_j$ controls the dimension of $\bftheta^{(j)}$. Based on the variogram analysis in Section \ref{secData}, we choose to use a smoothness ($\nu$) which is constant across segments and fixed to be $\nu = 0.5$ (true for all partitions), corresponding to an exponential correlation structure in (\ref{maternMod}). Otherwise,
\begin{equation} \label{theta_Prior}
p(\bftheta^{(j)} |\mcP_j) = \prod_{k=1}^{K_j} p( \tau_{jk}^2)  p( \sigma_{jk}^2)  p( \phi^{(1)}_{jk} )  p( \phi^{(2)}_{jk} )  p( \eta_{jk})
\end{equation}
(the conditioning on $\mcP_j$ on the right hand side is suppressed). Each component of (\ref{theta_Prior}) is noninformative and proper: for $k=1, \dots, K_j$, 
\begin{equation*} 
\begin{array} {c}
p(\tau_{jk}^2) = \text{Uniform}(0, 100), \hskip3ex
p(\sigma_{jk}^2) = \text{Uniform}(0, 100), \\[1ex]
p(\phi^{(1)}_{jk}) = \text{Uniform}(0, \sqrt{2}), \hskip3ex
p(\phi^{(2)}_{jk}) = \text{Uniform}(0, \sqrt{2}), \\[1ex]
p(\eta_{jk}) = \text{Uniform}[0, \pi/2]. 
\end{array}
\end{equation*}
Uniform priors for the variance parameters $\tau^2$ and $\sigma^2$ are used in place of more traditional conjugate inverse-Gamma priors to avoid prior bias in the case where an individual segment contains a small number of observations (following \citealp{Gelman2006}). For this analysis, the longitude/latitude coordinates of locations within each cluster were rescaled to lie in $[0,1] \times [0,1]$ to improve mixing of the MCMC; therefore, the upper limits for the priors on the anisotropy matrix eigenvalues, $\phi^{(1)}_{jk}$ and $\phi^{(2)}_{jk}$ (which correspond to squared ranges of dependence) are set to $\sqrt{2}$, which is the maximum distance on $[0,1] \times [0,1]$. The reasoning behind this choice is that the squared spatial range of the process within a segment is not expected to exceed the size of the segment. The limits on the prior for $\eta_{jk}$ are set to ensure identifiability (following, e.g., \citealp{Katzfuss2013}). 

\subsection{Model fitting and MCMC} \label{subsecMCMC}

Conditional on partition $\mcP_j$, the posterior distribution is
\begin{equation} \label{condPost}
p(\mu_j, \bftheta^{(j)} | \mcP_j, {\bf Z=z}) \propto p({\bf Z}| \mu_j, \bftheta^{(j)}, \mcP_j) \cdot p(\mu_j |\mcP_j) \cdot p(\bftheta^{(j)} |\mcP_j),
\end{equation}
which combines (\ref{likelihood}) and priors defined in Section \ref{subsecCond}. As usual, (\ref{condPost}) is not available in closed form, and we must resort to Markov chain Monte Carlo (MCMC) methods. Posterior samples from (\ref{condPost}) are generated using the {\tt nimble} package for $\R$ (\citealp{nimble_jcgs}): the MCMC for $\mcP_j$ is run for 20,000 total iterations with 10,000 iterations discarded as burn-in. A univariate Metropolis Hastings sampling step is conducted for the overall mean ($\mu_j$), and adaptive random walk samplers are conducted for the segment-specific variance/covariance parameters ($[\tau^2_{jk}, \sigma^2_{jk}, \phi^{(1)}_{jk}, \phi^{(2)}_{jk}, \eta_{jk}]$, sampled as a block).  

\subsection{Model-averaged posterior prediction} \label{subsecBMA}

Recall from Section \ref{secIntroduction} that our primary goal for using this model is prediction. Define a collection of locations $\{ \bfs^*_1, \dots, \bfs^*_m\} \subset \mathcal{D}$ for which we would like to obtain predictions of the corresponding (log) SOC values ${\bf Z}^* = ( Z(\bfs^*_1), \dots, Z(\bfs^*_m) )$. While the statistical model outlined in Section \ref{subsecCond} is conditional on a single partition $\mcP_j$, using a Bayesian framework we can average over the partitions $\{ \mcP_j: j = 1, \dots, 8 \}$ from Section \ref{secCovtPar} to obtain the full posterior predictive distribution for ${\bf Z}^*$ conditional on observed ${\bf Z=z}$:
\begin{equation} \label{postPred}
p({\bf Z}^* | {\bf Z=z} ) = \sum_{j=1}^{8} \int \int p({\bf Z}^*, \mu_j, \bftheta^{(j)}, \mcP_j | {\bf Z=z}) d\mu_j d\bftheta^{(j)}.
\end{equation}
The properties of conditional probabilities allow us to re-write (\ref{postPred}) as
\begin{equation} \label{postPred2}
p({\bf Z}^* | {\bf Z=z} ) = \sum_{j=1}^{8} \int \int p({\bf Z}^* | \mu_j, \bftheta^{(j)}, \mcP_j, {\bf Z=z}) \hskip15ex
\end{equation}
\[
\hskip25ex \times \>\> p(\mu_j, \bftheta^{(j)} | \mcP_j, {\bf Z=z}) \times p(\mcP_j | {\bf Z=z}) d\mu_j d\bftheta^{(j)}.
\]
This factorization of (\ref{postPred}) greatly simplifies posterior prediction, as follows. Since we are using a Gaussian process model, $p({\bf Z}^* | \mu_j, \bftheta^{(j)}, \mcP_j, {\bf Z=z})$ is multivariate Gaussian: with 
\begin{equation*}
\left[ \begin{array}{c} {\bf Z} \\ {\bf Z^*} \end{array} \Bigg| \hskip1ex \mu_j, \bftheta^{(j)}, \mcP_j \hskip1ex \right] \sim {N}_{n+m} \left( \mu_j {\bf 1}_{n+m} ,  \left[ \begin{array}{cc} {\bf V}_j + {\bf \Omega}_j & {\bf \Omega}_j^{\bf ZZ^*} \\  {\bf \Omega}_j^{\bf Z^*Z} & {\bf V}_j^* + {\bf \Omega}^*_j  \end{array} \right] \right),
\end{equation*}
where ${\bf \Omega}_j^{\bf Z^*Z} \equiv \text{Cov}_j({\bf Z^*, Z})$, it follows that
\begin{equation} \label{Zstar}
p({\bf Z}^* | \mu_j, \bftheta^{(j)}, \mcP_j, {\bf Z=z}) = {N}_m \left(\boldsymbol{\mu}^{(j)}_{\bf Z^*|z} , {\bf \Sigma}^{(j)}_{\bf Z^*|z} \right),
\end{equation}
where 
\[
\boldsymbol{\mu}^{(j)}_{\bf Z^*|z} =  \mu_j {\bf 1}_{m} +  {\bf \Omega}_j^{\bf Z^*Z}  ({\bf V}_j + {\bf \Omega}_j)^{-1} ({\bf z} - \mu_j {\bf 1}_{n}),
\]
\[
{\bf \Sigma}^{(j)}_{\bf Z^*|z} = ( {\bf V}_j^* + {\bf \Omega}_j^* ) - {\bf \Omega}_j^{\bf Z^*Z} ({\bf V}_j + {\bf \Omega}_j)^{-1}  {\bf \Omega}_j^{\bf ZZ^*}.
\]
The next component in (\ref{postPred2}) is the conditional posterior distribution (\ref{condPost}), while the final component of (\ref{postPred2}) is the posterior probability of each partition $\mcP_j$ conditional on the data. From Bayes' Theorem, the latter is 
\begin{equation} \label{modelPost}
p(\mcP_j | {\bf Z=z}) = \frac{ p({\bf Z} | \mcP_j)  p(\mcP_j) }{ \sum_{i=1}^{8} p({\bf Z} | \mcP_i)  p(\mcP_i) } = \frac{ p({\bf Z} | \mcP_j) }{ \sum_{i=1}^{8} p({\bf Z} | \mcP_i) }.
\end{equation}
The last equality follows from our use of a uniform prior over the different partitions, i.e., $p(\mcP_j) = 1/8$, $j = 1, \dots, 8$, which is appropriate in this setting because the partitions are defined by the covariates and not SOC. In any case, the important quantity here is the marginal likelihood
\begin{equation} \label{margLik}
p({\bf Z} | \mcP_j) = \int \int p({\bf Z} | \mu_j, \bftheta^{(j)}, \mcP_j)  p(\mu_j, \bftheta^{(j)} | \mcP_j) d\mu_j d\bftheta^{(j)}.
\end{equation}
Estimation of the marginal likelihood is a well-known problem in Bayesian analysis (for more details on our approach, see Appendix \ref{appdxMargLikEstimation}).

Combining all of the above, (\ref{postPred2}) suggests the following algorithm to sample from $p({\bf Z}^* | {\bf Z=z} )$:

\begin{enumerate}
\item Draw $\mcP_j $ according to the $p(\mcP_j | {\bf Z=z}), j = 1,\dots, 8$. 
\item Draw $(\mu_j^*, \bftheta^{(j)}_*)$ from $p(\mu_j, \bftheta^{(j)} | \mcP = \mcP_j, {\bf Z=z})$, which is the posterior distribution for the mean and covariance parameters conditional on $\mcP_j$ (the sampling of which is described in Section \ref{subsecMCMC}).
\item Draw ${\bf Z}^*$ from $p({\bf Z}^* | \mu_j = \mu_j^*, \bftheta^{(j)} = \bftheta^{(j)}_*, \mcP = \mcP_j, {\bf Z=z})$ as defined in (\ref{Zstar}).
\end{enumerate}

\section{Results} \label{secResults}
\subsection{Posterior prediction maps} \label{subsecPredMap}
Before showing the prediction maps, recall that the sampling algorithm requires estimates of the marginal likelihood. A variety of methods for estimating the marginal likelihoods (\ref{margLik}), outlined in Appendix \ref{appdxMargLikEstimation}, were applied to the {eight} partitions. The resulting log marginal likelihood estimates (scaled to have mean zero and variance one) and normalized probabilities (\ref{modelPost}) are plotted in Figure \ref{postModelProbs} of Appendix \ref{appB3}. {The scaled log marginal likelihood estimates show agreement between all of the estimators except BICM, which we henceforth exclude from consideration. Otherwise, on the probability scale, the estimators collectively give non-zero weight to the same three or four partitions (4, 6, 7, and 8) and none of the methods suggest that the model probabilities should be evenly distributed across all eight partitions. Given this relative agreement and for simplicity, we decided to use the HM estimator for generating posterior predictions of log SOC.}

\begin{figure}[!t]
\begin{center}
\includegraphics[width=\textwidth]{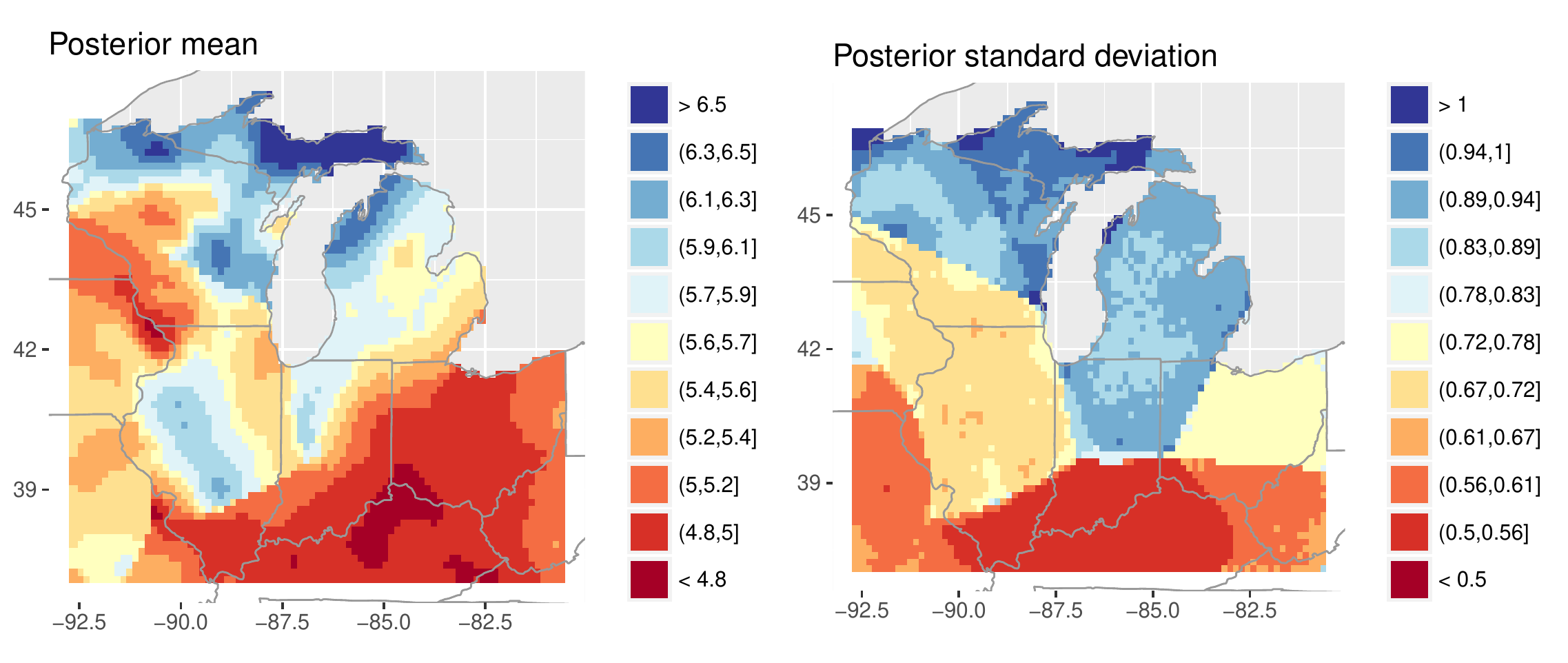}
\caption{Model-averaged posterior mean predictions (left) and corresponding posterior standard deviations (right) for log SOC (units = log Mg/ha to 100cm).}
\label{maPreds}
\end{center}
\end{figure}

The resulting model-averaged posterior prediction maps and corresponding standard deviations for the Great Lakes region are shown in Figure \ref{maPreds}. {By averaging over the partitions, which individually consist of independent segments, we are able to obtain a relatively ``smooth'' posterior mean surface, although artifacts of the individual segments are visible in the posterior standard deviation map. While the standard deviation map is not smooth, we note that this non-smoothness is what the SOC data select as the best model fits (based on the marginal likelihood estimates), and other marginal likelihood estimators give qualitatively similar results (see Figure 8). Regardless, the standard deviation map characterizes how the variability in predictions of log SOC varies over space; see Section \ref{subsecImpl} for further discussion.} {Comparing both the predicted (log) SOC surface obtained using our non-stationary Gaussian process model (Figure \ref{maPreds}) as well as its standard deviation map to those obtained using other off-the-shelf methods (Figure \ref{otherPreds}) we can conclude that SOC is much more flexibly modeled with our proposed nonstationary model, as the prediction variances are better described by our model.}

{Additionally}, {another} major benefit of using our nonstationary method with covariate-partitioning {for the SOC data} is the significant reduction in computational time relative to fitting a stationary Bayesian spatial model (as well as other related Bayesian models). Recall that the {eight} candidate partitions are generated ahead of time, and then, for each partition, the nonstationary models can be fit at the same time in parallel. For a dataset of size  $n \approx 800$ observations, the maximum time required for an individual partition was {7} minutes (for 20,000 MCMC iterations), and the average time across all partitions was just {3.4} minutes (see Table \ref{modelComp}). A traditional stationary model (e.g., SGP; see Section \ref{subsecOffshelf}) took approximately 20 minutes (computational times given are for an Intel Core i7 3.1 GHz machine with 16 GB memory). This major improvement in computational time is due to the fact that the likelihood conditional on a partition is the product of independent multivariate Gaussian likelihoods (see Section \ref{subsecCond}).
  
\subsection{Model comparison} \label{subsecModelComp} 

{Besides the qualitative comparison of the predictive SOC maps generated by our} model-averaged, covariate-partitioned nonstationary model {(henceforth NSGP)} {and the other off-the-shelf models, we also compare} {the predictive performance of the various models quantitatively. To this goal, we first}
use 10-fold cross validation in which the full data set ($n = 790$) is split into ten equally sized test data sets with $m=79$ observations. For each subset, we fit all models using the other nine subsets as training data and obtain predictions at the test locations. The continuous rank probability score ($\CRPS$; \citealp{PropScoring}) is used to evaluate the sharpness and calibration of predictions for the held-out data. \cite{Kruger2016} outline several methods for estimating the $\CRPS$ for individual predictions based on the output from MCMC algorithms (i.e., $\widehat{\CRPS}_t$ for $t = 1, \dots, m$). Mean scores are calculated for the test set using
$
\widehat{\CRPS} = m^{-1}  \sum_{t=1}^m \widehat{\CRPS}_t
$  
and compared across models. Since the conditional predictive cumulative distribution functions (conditional on the true parameter state, i.e., the likelihood) are not available for BART and TGP, we use what \cite{Kruger2016} call the empirical CDF method (ECDF) for calculating $\CRPS$ for each model. ECDF is based on samples drawn from the posterior predictive distribution and is a consistent approximation for the $\CRPS$ for every predictive distribution with finite mean (\citealp{Kruger2016}). The ECDF estimate of $\CRPS$ is calculated using the {\tt scoringRules} package for \texttt{R} (\citealp{scoringRules}).

{For the 10-fold cross validation, the CRPS for each fold is the average of univariate CRPS across the test data. Given that the our approach targets second-order properties of log SOC, we would also like to evaluate the models with respect to a multivariate criteria, e.g., the energy score. However, in other work we have found it difficult to implement the energy score, because deviations in the energy score are challenging to diagnose. Instead, we evaluate the models' predictive capability for spatial \textit{averages} of log SOC. 
{We choose to work with spatial averages because while spatial averages are still univariate variables (and can thus be evaluated using the aforementioned methods), by being functionals of the joint predictive distribution, their second moment depends on the covariance structure of the (log) SOC process. Additionally, spatial averages are more compelling and useful quantities to examine from a soil management perspective.} 
To this end, we fit all models to two additional groups of holdout sets, block and circular. Each block holdout set (12 total) creates a test data set out of a contiguous $\approx 3.3^\circ$ longitude by $\approx 1.7^\circ$ latitude box (the holdout sets range in size from 29 to 79 locations; see Figure \ref{blockHdt} in the Appendix). Each circular holdout set (10 total) creates a test data set using the 29 nearest neighbors of a randomly selected station (so that the holdout sets consist of 29 + 1 = 30 stations; see Figure \ref{circHdt} in the Appendix). For each holdout set (block and circular), the other sets are used as training data to predict the spatial average corresponding to the test data locations. Then, we calculate the univariate CRPS for the spatial average, again using the {\tt scoringRules} package as described previously.}

\begin{table}[!t]
\caption{ Mean continuous rank probability scores (averaged over holdout sets) for each group of holdout sets. The best model for each holdout set is in bold.}
\begin{center}
\begin{tabular}{|c||c|c|c|c|c|}
\hline
\textbf{Holdout sets} 	& \textbf{NSGP} & \textbf{SGP} & \textbf{BART}$^\dagger$  & \textbf{TGP} & \textbf{PP} \\ \hline\hline 
10-fold 			& \textbf{0.3912} & 0.3959 & 0.4067 & 0.3988 & 0.3978\\ \hline
Block			& 0.1674 & 0.1851 & 0.1679 & \textbf{0.1357} & 0.1883\\ \hline
Circular 			& \textbf{0.1784} & 0.2293 & 0.1505 & 0.1889 & 0.2227\\ \hline
\end{tabular}
\end{center}
\label{outCV_table}
\begin{flushleft}
\vskip-2ex
{\scriptsize $^\dagger$BART is excluded from consideration based on its unrealistic prediction map.}\\
\end{flushleft}
\end{table}%

{The CRPS for each model in Table \ref{modelComp} and each group of holdout sets are summarized in Table \ref{outCV_table}. Our covariate segmentation nonstationary model yields the best CRPS when averaging the univariate CRPS over locations in each fold, but the improvement relative to the other models is modest. The two nonstationary models (NSGP and TGP) perform well for the block holdout sets, although TGP maintains a clear advantage over NSGP. This is not completely surprising, since TGP (which uses rectangular partitions of the domain) is expected to perform well for rectangular holdout sets. When looking at the circular holdout sets, NSGP is preferred to TGP, which again is expected since NSGP uses non-rectangular partitions of the domain. BART outperforms both TGP and NSGP for the circular holdout sets, although recall that we previously decided to exclude BART from consideration based on its unrealistic prediction map.}

One limitation of using CRPS to compare models is that it is difficult to evaluate the relative improvement for one model versus another. Clearly, NSGP yields only modest (at best) improvements in CRPS relative to the other fitted models; however, we note that other papers that focus on non-stationary modeling (e.g., \citealp{Paciorek2006}; \citealp{Fuglstad2015}) also find very small improvements in out-of-sample evaluation criteria relative to stationary models, even when exploratory analyses indicate the presence of nonstationarity in the data. \cite{Fuglstad2015} note that, in their experience, ``non-stationary models do not lead to much difference in the predicted values, [but] that the differences are found in the prediction variances'' -- this is also noted by \cite{schmidt11} and \cite{ViannaNeto}, and is certainly true for our application as well. As in \cite{Fuglstad2015}, we would like to point out that whether or not we have improved the CRPS is not the only question worth asking: in this case, we argue that our more flexible characterization of the prediction variances yields increased insight into the spatial distribution of SOC (we expound upon this further in the next section).

\subsection{Implications for decision making} \label{subsecImpl}

Gridded prediction maps and corresponding standard deviations such as those shown in Figures \ref{otherPreds} and \ref{maPreds} are extremely important to soil scientists, and (as mentioned in Section \ref{secIntroduction}) are used for a wide variety of purposes, including benchmarking mechanistic models of soil carbon (e.g., \citealp{ToddBrown2014}), identifying target areas for soil restoration projects (e.g., \citealp{Ryals2014}; \citealp{Ryals2015}), and informing carbon sequestration projects that seek to determine the limits of soils' carbon storing capacity (e.g., \citealp{Angers2011}; \citealp{Wiesmeier2014}). For each of these purposes, it is very important to account for spatial variation in the uncertainty of the resulting predictions: as discussed in Section \ref{subsecModelComp}, note that the NSGP model can account for this (e.g., right panel of Figure \ref{maPreds}) while the stationary (or approximately stationary) models cannot (e.g., bottom row of Figure \ref{otherPreds}). Thus, while the posterior mean predictions look quite similar (for example, NSGP, SGP, and PP) and 
{the quantitative cross validation results do not yield overwhelming evidence in support of the nonstationary model,} 
the clear differences in standard deviations are a strong argument for using NSGP. 

\begin{figure}[!t]
\begin{center}
\includegraphics[width=\textwidth]{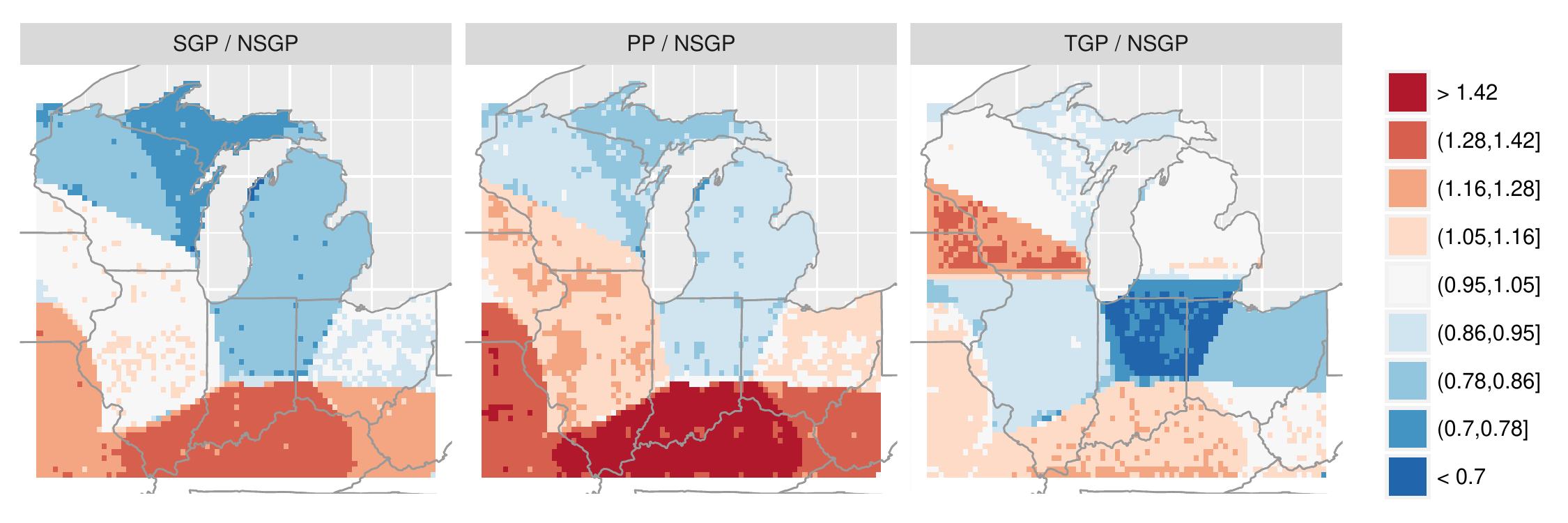}
\caption{Comparison of the posterior standard deviations across SGP, PP, TGP, and NSGP. The plotted colors represent the ratio of standard deviations for SGP (left), PP (center), and TGP (right) divided by the standard deviations for NSGP. Red areas correspond to regions where NS-GP provides more precise predictions relative to the other models; blue areas represent regions where the other models likely underestimate the standard deviations.}
\label{stdErrRatio}
\end{center}
\end{figure}

To emphasize this point, we plot the ratio of standard deviations in Figure \ref{stdErrRatio}, specifically showing the posterior standard deviations for SGP, PP, and TGP divided by the corresponding quantity for NSGP. In this plot, red areas correspond to regions where NSGP provides much more precise predictions relative to the other models; blue areas represent regions where posterior standard deviations from the other models are smaller. Using the results from the nonstationary model, a practitioner seeking to identify target areas for a new soil restoration project may want to focus their efforts in the Ohio River Valley (Kentucky, southern Indiana, and southern Illinois), as this is an area where both the soil carbon predictions are low and the uncertainty associated with these predictions is small. Alternatively, while soil carbon appears to be low along the Mississippi River in Wisconsin, the uncertainty in these measurements is much larger. Neither SGP nor PP provide this information.

\section{Discussion} \label{secDiscussion}

In this paper, we have proposed a novel method for obtaining predictions of soil organic carbon at unobserved locations using a covariate-driven nonstationary spatial Gaussian process model. Our approach uses covariate partitioning to divide the geographic space according to the within-segment distribution of the covariates (here, land use-land cover and drainage class), so that the covariates indirectly inform the modeled first- and second-order properties of SOC as well as the resulting predictions. This approach extends existing covariate-driven nonstationary approaches in that the information a covariate provides on SOC can be used for prediction even if the covariate is not fully observed over the spatial domain. Bayesian model averaging accounts for uncertainty in the partitioning, and while our approach yields only modest improvements in terms of out-of-sample evaluation criteria, it provides a more appropriate characterization of the spatial variation in prediction uncertainty. Furthermore, while many nonstationary spatial approaches are computationally intensive, our approach results in computational times that are significantly faster than the time required to fit a corresponding second-order stationary model.

Of course, there are limitations to our nonstationary model. For example, the marginal likelihood estimation seems to be fairly sensitive to individual likelihood values, resulting in a few non-zero and many nearly zero posterior model probabilities. Regardless of which marginal likelihood method was used (see Appendix \ref{appdxMargLikEstimation}), the estimates always put nearly all the weight on a single model. In a general setting, this may not be problematic; however, in our approach, we rely on model averaging to produce a scientifically meaningful (i.e., smooth) fitted surface, and we are well aware that the individual models, which specify independence across segments, are not on their own appropriate for modeling an environmental process like SOC. In light of the variability in the marginal likelihood estimates, we might also consider a {second} approach to model averaging, {which} could involve identifying a subset of the partitions that are in some sense ``good'' (as indicated by the data) and uniformly weighting over the reduced set. The resulting surface would likely be more scientifically meaningful while also using the data to indicate which of the partitions provide a better fit to the observed SOC.

Finally, an additional extension that might address the non-smooth mean predictions in Figure \ref{maPreds} would be to use different weighting functions for the partition-specific likelihood in (\ref{likelihood}). Recall that we use indicator weight functions; more generally, we might use weight functions without sharp boundaries. In this case, the transition between segments for a particular partition would be smoother, resulting in a more realistic specification for the individual models. In fact, we implemented such an approach, using the bivariate Gaussian densities from (\ref{eq:clusters}) as the unnormalized weights combined with the nearest neighbor Gaussian process likelihood (NNGP; \citealp{Datta2016}) to speed up the computation. Unfortunately, such an approach did not improve the CRPS criteria relative to NSGP with indicator weights, and the computational times for fitting the individual partitions increased to as much as 2 hours (even with the NNGP likelihood; without NNGP the computational times were much longer, exceeding 10 hours for an individual partition). In general, the primary benefit of using indicator weight functions is computational: the product nature of (\ref{likelihood}) greatly reduces the computational time for fitting each model. The decision to use a different weight function should be balanced with the computational limitations of model fitting for a general data set.

\section*{Acknowledgements}
The authors would like to acknowledge Dr. Katerina Georgiou, a soil scientist at the Lawrence Berkeley National Laboratory, for helpful discussion regarding soil organic carbon. This work was supported in part by the Statistical Methods for Atmospheric and Oceanic Sciences (STATMOS) research network (NSF-DMS awards 1106862, 1106974, and 1107046). C. A. Calder was also partially supported by NSF-DMS award 1209161.

\bibliographystyle{imsart-nameyear} \bibliography{MDRResearch}

\appendix

\section{Marginal likelihood estimation} \label{appdxMargLikEstimation}

\subsection{Importance sampling techniques}

The original citations for what follows are \cite{rosenkranz1992}, \cite{newton1994}, \cite{carlin1995}; however, a helpful summary is provided by Raftery in Chapter 10 (section 3) of \cite{MCMCinPractice}. 

Recall from (\ref{margLik}) that the marginal likelihood for an individual partition is $p({\bf Z} | \mcP_j) = \int \int p({\bf Z} | \mu_j, \bftheta^{(j)}, \mcP_j)  p(\mu_j, \bftheta^{(j)} | \mcP_j) d\mu_j d\bftheta^{(j)}$. For notational simplicity, suppress the implicit dependence on $\mcP_j$ and in an abuse of notation rewrite $\bftheta = (\mu_j, \bftheta^{(j)})$. What we want is $p({\bf Z}) = \int L(\bftheta) \pi(\bftheta) d\bftheta$,
where $\pi(\bftheta)$ is the prior distribution and $L(\bftheta) \equiv p({\bf Z} | \bftheta)$ is the likelihood. Furthermore, define
$
|| X ||_h = T^{-1} \sum_{t=1}^T X(\bftheta^{(t)}),
$
where $\{ \bftheta^{(t)}: t= 1,\dots,T\}$ are samples from the density $h(\bftheta)/\int h(\boldsymbol{\phi}) d\boldsymbol{\phi}$. Importance sampling supposes we can sample $\bftheta$ from $c g(\bftheta)$, where $c^{-1} = \int g(\bftheta)d\bftheta$; then
\begin{equation*} \label{IS}
p({\bf Z}) = \int L(\bftheta)  \pi(\bftheta) d\bftheta = \int L(\bftheta)  \left[ \frac{\pi(\bftheta)}{c g(\bftheta)} \right]  c g(\bftheta)d\bftheta.
\end{equation*}
Given $\{ \bftheta^{(t)}: t= 1,\dots,T\}$ from $c \cdot g(\bftheta)$, then
\begin{equation} \label{ISestimate}
\widehat{p}({\bf Z}) = \frac{1}{T} \sum_{t=1}^T \frac{ L( \bftheta^{(t)}) \pi(\bftheta^{(t)}) }{ c g(\bftheta^{(t)}) } \equiv \left|\left| \frac{L\pi}{cg} \right|\right|_g.
\end{equation}
If $c$ is unknown, it can be estimated similarly by $\widehat{c} = ||\pi/g||_g$, so that (\ref{ISestimate}) becomes
$\widehat{p}({\bf Z}) =\left|\left| L\pi /g \right|\right|_g / \left|\left| \pi/g \right|\right|_g $. 

As a simple option, \cite{newton1994} suggest using $g = L\pi$ (i.e., the posterior $p(\bftheta|{\bf Z})$) since posterior simulation already provides samples from this choice of $g$. In this case,
\begin{equation} \label{phat1}
\widehat{p}_{HM}({\bf Z}) = \frac{1}{ \left|\left| 1/L \right|\right|_{\text{post}} }
\end{equation}
where ``post'' indicates that the samples of $\bftheta$ are from the posterior (the ``HM'' subscript indicates that this is the harmonic mean of the likelihood values). This estimator converges almost surely to $p({\bf Z})$ as $T\rightarrow \infty$, but does not satisfy a Gaussian CLT since the variance of $1/\widehat{p}_{HM}({\bf Z})$ is usually non-finite. However, \cite{rosenkranz1992} concludes that this estimator is a good choice because it is easy to compute and performs well when $T$ is large ($\geq 5,000$), a result reiterated by \cite{carlin1995}.

More generally, $g$ can be $g_\delta(\bftheta) = \delta \cdot \pi(\bftheta) + [1-\delta] \cdot p(\bftheta|{\bf Z})$ (\citealp{newton1994}), where $0 < \delta < 1$ and $p(\bftheta|{\bf Z})$ is the posterior distribution, because the mixture of the prior and posterior yields high efficiency and satisfies a Gaussian CLT. In practice, one can simulate $T$ values of $\bftheta$ from the posterior and suppose that an additional $\delta T/(1-\delta)$ values are drawn from the prior, each of which has a likelihood $L(\bftheta)$ equal to its expected value $p({\bf Z})$. Then, the estimate $\widehat{p}_\delta({\bf Z})$ can be obtained by finding the solution $x$ to the equation
\[
x = \frac{ \frac{\delta T}{1-\delta} + \sum_{t=1}^T \frac{L(\bftheta^{(t)})}{\delta x + (1-\delta)L(\bftheta^{(t)})} }{ \frac{\delta T}{x(1-\delta)}  + \sum_{t=1}^T \big[\delta x + (1-\delta)L(\bftheta^{(t)})\big]^{-1} }
\]
where the $\{\bftheta^{(t)}: t= 1, \dots, T\}$ are the posterior samples (page 169, Chapter 10, \citealp{MCMCinPractice}), which can be done using an iterative Newton-Raphson scheme. \cite{rosenkranz1992} found that large values of $\delta$ (i.e., near 1) yield the best performance.

\subsection{Monte Carlo approximations to AIC and BIC}

\cite{Raftery2007} present a variety of methods for estimating the marginal likelihood using only the likelihoods from posterior simulation output. In fact, their work specifically aims to improve the instability (i.e., infinite asymptotic variance) of the harmonic mean estimator (\ref{phat1}).

One of the approaches in \cite{Raftery2007} is based on the fact that the posterior distribution of the log likelihood approximately follows a gamma distribution. As such, they note that the corresponding estimate of the marginal likelihood has an interesting similarity to the Bayesian information criterion (BIC), which yields a Monte Carlo approximation to the log marginal likelihood (henceforth BICM)
\begin{equation*} \label{BICM}
\log \widehat{p}_{\BICM}({\bf Z}) = \overline{\mathcal{L}} - s^2_{\mathcal{L}} (\log n - 1),
\end{equation*}
where $\overline{\mathcal{L}}$ and $s^2_{\mathcal{L}}$ are the sample mean and variance of the log likelihoods from the posterior samples and $n$ is the sample size. Similarly, a posterior simulation-based version of Akaike's information criterion (AIC) can be used to approximate the log marginal likelihood (henceforth AICM)
\begin{equation*} \label{AICM}
\log \widehat{p}_{\AICM}({\bf Z}) = 2(\overline{\mathcal{L}} - s^2_{\mathcal{L}}).
\end{equation*}

\subsection{Results} \label{appB3}

\begin{figure}[!h]
\begin{center}
\includegraphics[width=\textwidth]{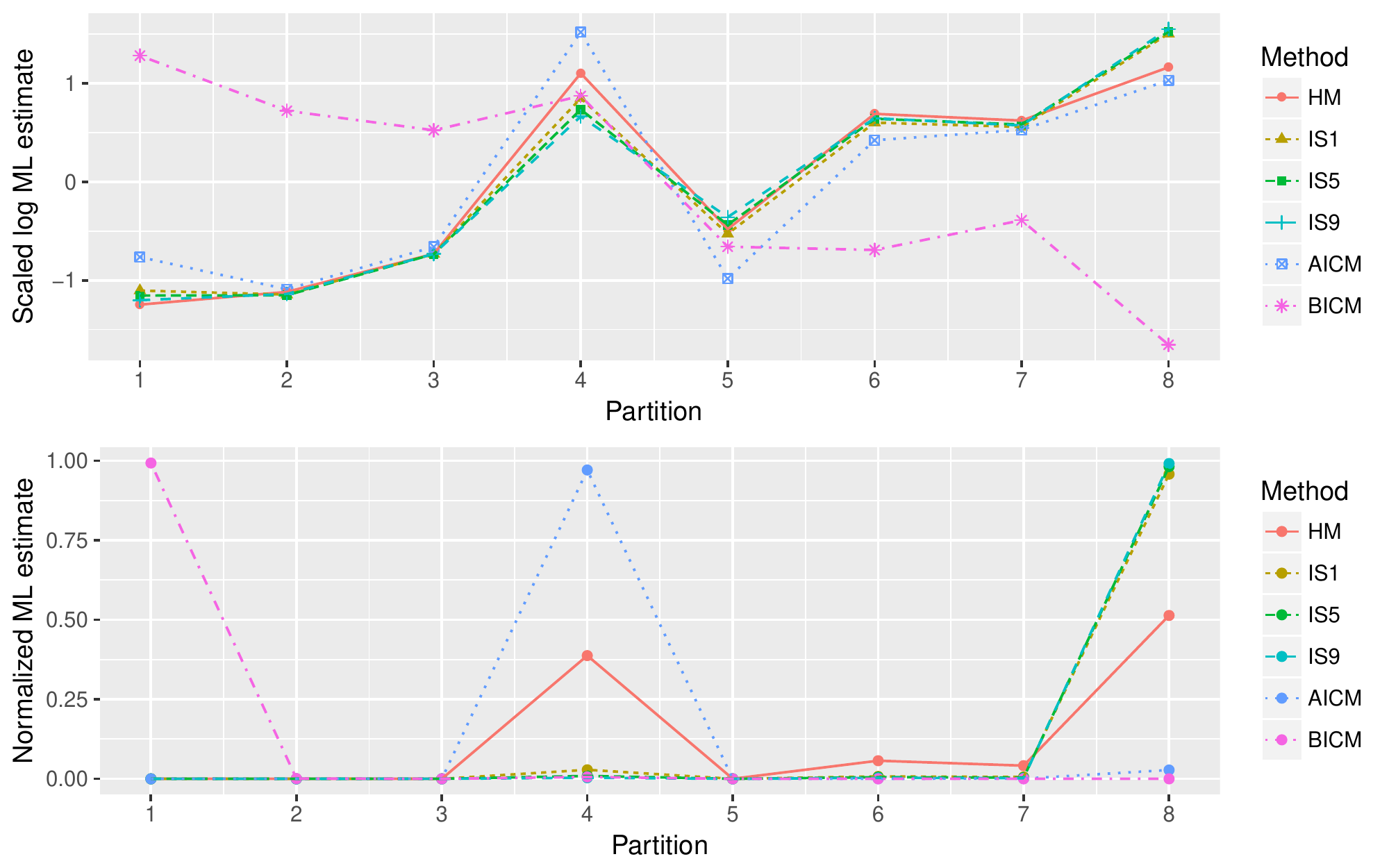}
\caption{Scaled log marginal likelihood (ML) estimates and normalized ML estimates (probabilities $p(\mcP|{\bf Z=z})$) of each partition for each of several ML estimation methods. AICM and BICM are Monte Carlo estimates of AIC and BIC, respectively (\citealp{Raftery2007}); HM is the harmonic mean estimator (\citealp{KassRaftery1995}). ``IS$x$'' corresponds to an importance sampling estimate where the proposal density is a convex combination of the prior and posterior, with $100(x/10)\%$ of the proposal samples coming from the prior and $100(1 - x/10)\%$ from the posterior (again see \citealp{KassRaftery1995}).}
\label{postModelProbs}
\end{center}
\end{figure}

Each of the approaches in the previous two sections were applied to the real data example. The results are shown in Figure \ref{postModelProbs}.

\clearpage
\section{Supplemental figures}
%

\begin{figure}[!h]
\begin{center}
\includegraphics[width=0.8\textwidth]{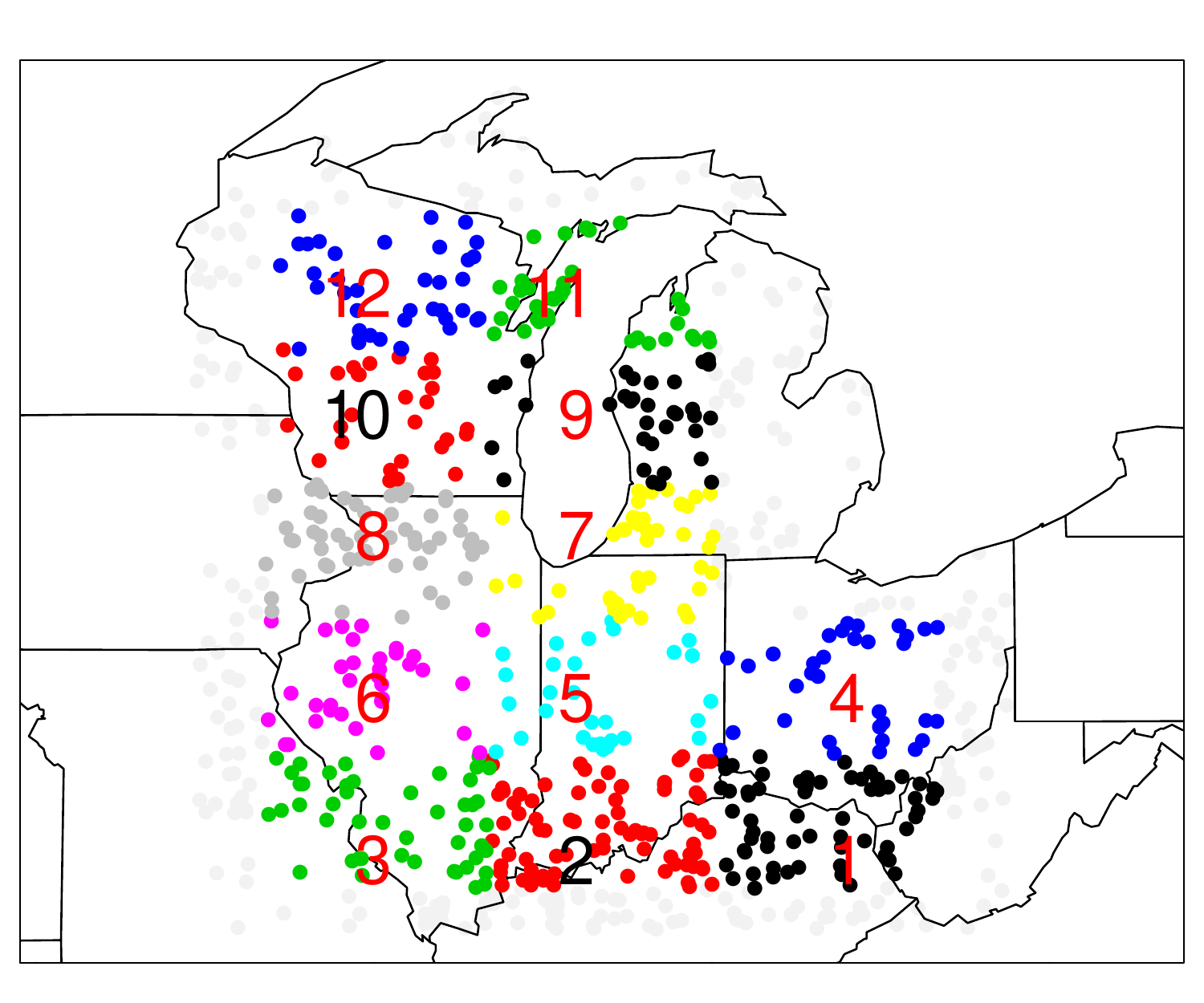}
\caption{The 12 block holdout sets.}
\label{blockHdt}
\end{center}
\end{figure}

\begin{figure}[!h]
\begin{center}
\includegraphics[width=\textwidth]{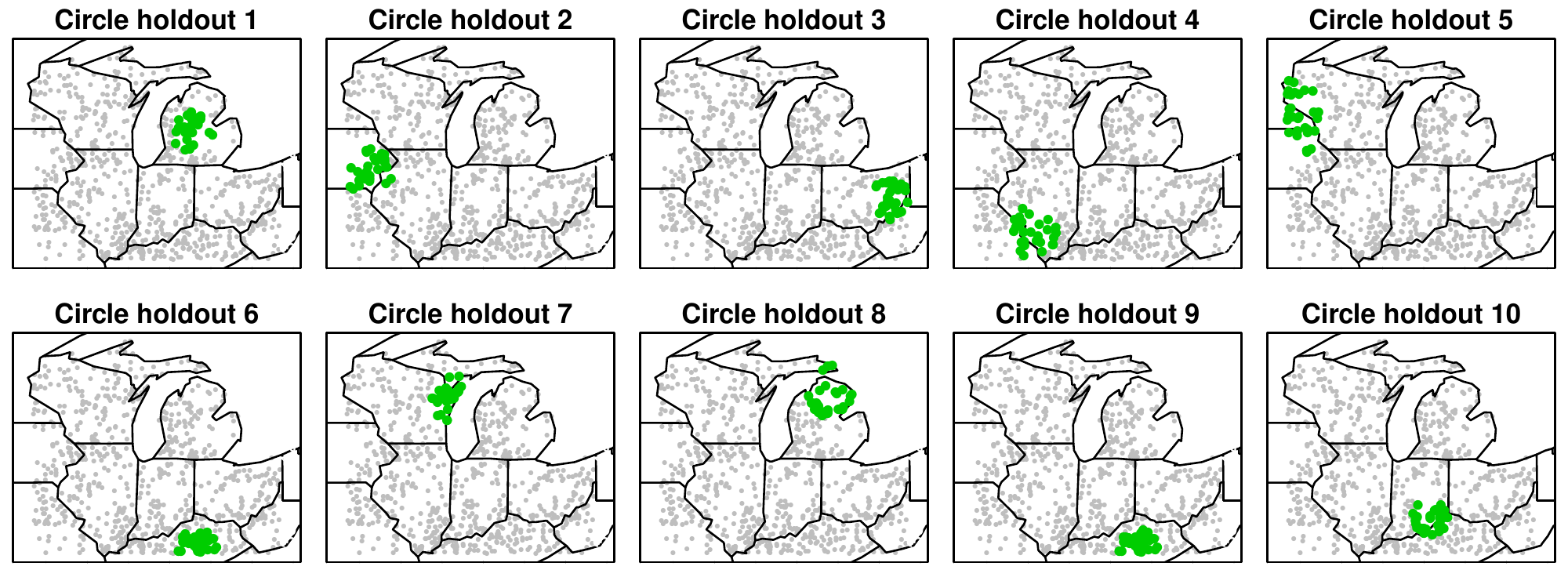}
\caption{The 10 circular holdout sets.}
\label{circHdt}
\end{center}
\end{figure}

\end{document}